# Earth's carbon deficit caused by early loss through irreversible sublimation

**J. Li[1,*], E.A. Bergin[2], G.A.Blake[3],  F.J.Ciesla[4], M.M. Hirschmann[5]**

[1]Department of Earth and Environmental Sciences, University of Michigan, Ann Arbor, MI 48109.

[2]Department of Astronomy, University of Michigan, Ann Arbor, MI 48109.

[3]Division of Geological & Planetary Sciences, California Institute of Technology, Pasadena, CA 91125.

[4]Department of Geophysical Sciences and Chicago Center for Cosmochemistry, University of Chicago, Chicago, IL 60637.

[5]Department of Earth and Environmental Sciences, University of Minnesota, Minneapolis, MN 55455.

**\*Correspondence to: jackieli@umich.edu**

## Abstract

Carbon is an essential element for life but its behavior during Earth's accretion is not well understood. Carbonaceous grains in meteoritic and cometary materials suggest that irreversible sublimation, and not condensation, governs carbon acquisition by terrestrial worlds. Through astronomical observations and modeling we show that the sublimation front of carbon carriers in the solar nebula, or the soot line, moved inward quickly so that carbon-rich ingredients would be available for accretion at 1 AU after the first million years. On the other hand, geological constraints firmly establish a severe carbon deficit in Earth, requiring the destruction of inherited carbonaceous organics in the majority of its building blocks. The carbon-poor nature of the Earth thus implies carbon loss in its precursor material through sublimation within the first million years.

## Introduction

Carbon provides an essential constituent of terrestrial life and plays a critical role in maintaining the Earth's habitable environment. Elucidating the processes of carbon acquisition by rocky planetary bodies is therefore crucial for understanding planetary habitability. Conventional condensation models for Earth formation typically assume chemical equilibrium and posit that elements were initially present in nebular gas and became available for accretion into solid bodies when the gases were sufficiently cool to precipitate (*1-2*).  As seen in Fig. 1, for refractory and moderately volatile lithophile elements (such as Mg, Si, Na, K), the degree of depletion in the bulk silicate Earth correlates with the element's half-mass condensation temperature (*1*), consistent with the condensation model. However, equilibrium condensation cannot explain the quantities and forms of carbon found in primitive chondrites and comets. At $10^{-3}$ bar pressure, major carbon carriers ($CO$, $CO_2$, $CH_4$) in the nebular gas do not condense at temperatures above 80 K, and therefore remain gaseous within tens of AU from the forming Sun (*1,3-4*). Consequently, chondrites and terrestrial worlds would have received no carbon. The presence of notable, although depleted, carbon in chondritic meteorites thus requires the creation or preservation of more refractory carbon-rich solids within the solar nebula than predicted by equilibrium condensation models.

Clear evidence for the survival of unprocessed pre-stellar grains has been found in the meteoritic record (*5*). These carbonaceous organics are not products of condensation from a hot atomized nebula, and therefore some or all must have been inherited from the interstellar medium (ISM) without ever being sublimated. However, the amount of carbon locked away as refractory solids in the ISM is far greater than carried by the most primitive and least processed meteorites (CI chondrites) that otherwise reflect solar composition (*6-7*). The presence of inherited carbon-rich solids at sub-ISM levels thus suggests that interstellar carbonaceous grains are partially destroyed in the inner solar system. Here we model the sublimation sequence and loss of carbon in the solar nebula. These results are then combined with derived upper bounds on the carbon content of the bulk Earth to investigate how nebular processes influence the acquisition of carbon by rocky planetary bodies.

## Results

The carbon content of solid aggregates in a protoplanetary disk depends on the extent of heating they experience, hence the sublimation sequence of carbon carriers as a function of nebular temperature rises to prominence as the governing process for carbon acquisition by terrestrial worlds. Astronomical observations show that approximately half of the cosmically-available carbon entered the protoplanetary disk as volatile ices and the other half as carbonaceous organic solids (*6*). As the disk warms up from 20 K, all the volatile carbon carriers sublimate by 120 K, followed by the conversion of major refractory carbon carriers into CO and other gases near a characteristic temperature of ~500 K (Table S1, Fig. S1). The sublimation sequence of carbon exhibits a "cliff" where dust grains in an accreting disk lose most of their carbon to gas within a narrow temperature range near 500 K (Fig. 1).

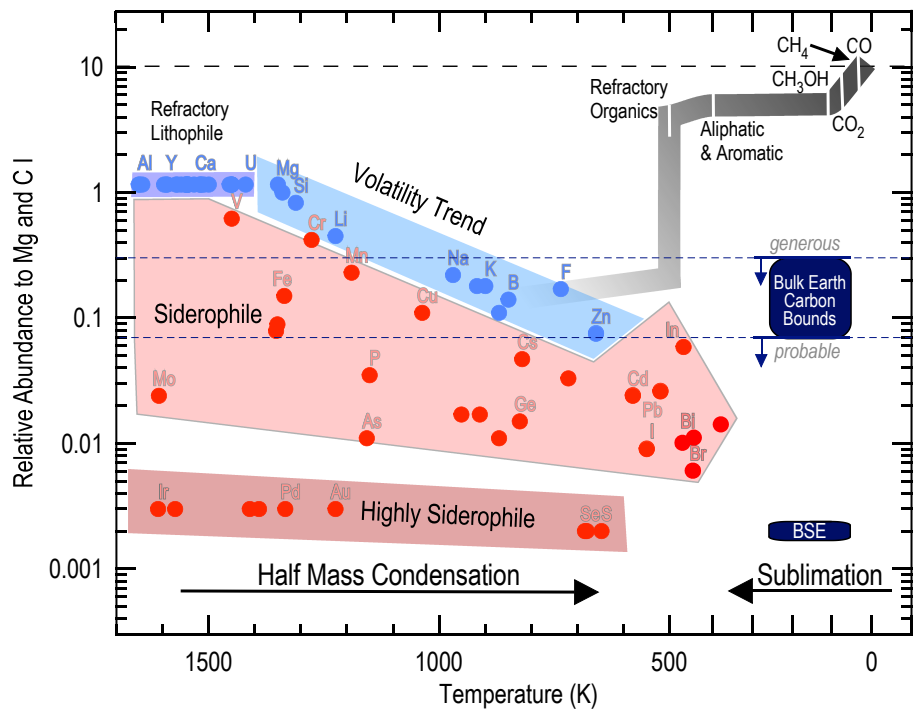

**Fig. 1 The sublimation sequence of carbon in the solar nebula.** An element's relative abundance to Mg and CI is the ratio of its relative abundance to Mg to that in CI, calculated as (C/Mg in Earth or solar) / (C/Mg in CI) in wt.%. On the high-temperature side, the volatility trend (blue shaded band) describes the relative abundances of lithophile elements (rock-loving, blue circles) in the bulk silicate Earth (mantle + crust) as a function of their half mass condensation temperatures (*1*). Siderophile elements (iron-loving, red circles) plot below the volatility trend, presumably due to preferential incorporation into the core. On the low-temperature side, the sublimation sequence of carbon (gray thick line) traces the falling relative abundance of condensed carbon in the solar nebula (excluding H and He) as the disk warms up (Table S4, Table S5). The abundance (relative to Mg and CI) of condensed carbon

starts at 9.48 (long-dashed line) and is reduced to 4.74 after carbon-carrying ices transform into gases. It falls precipitously by more than one order of magnitude when the nebular temperature reaches ~500 K. The maximum amount of carbon in the bulk Earth (large dark blue box), represented by a generous upper bound of 1.7±0.2 wt.% carbon (0.30±0.03 relative to Mg and Cl) and a probable bound of 0.4±0.2 wt.% carbon (0.07±0.04 relative to Mg and Cl), corresponds to a maximum fraction of 1-7% carbon-rich source material that survived sublimation (Table S6). The bulk silicate Earth (BSE, small dark blue box) with 140±40 ppm carbon by weight (1.7±0.5x10⁻³ relative to Mg and Cl) plots well below the upper bounds, possibly due to sequestration of carbon by the core.

The division between the stability fields of solid and gas carbon carriers corresponds to the "soot line", a term coined to describe the location where the irreversible destruction of pre-solar polycyclic hydrocarbons via thermally-driven reactions in the planet-forming region of disks occurred (*8*). In the earliest phases of star formation, the soot line migrates with time as the pressure and temperature of the disk evolve (Fig. 2). Disk observations suggest an order of magnitude variation in the accretion rate at a given time, but there is a clear general decline of accretion rate with age (*9*). Very early after the Sun's birth, when high accretion rates from the solar nebula onto the Sun provided a source of disk heating, $T_{soot}$ of 500 K may have been reached out to tens of AU near the disk midplane. As the accretion rate diminished and stellar irradiation became the dominant source of heating in the disk, the soot line then migrated inward to eventually cross 1 AU and reside interior to Earth's current orbit.

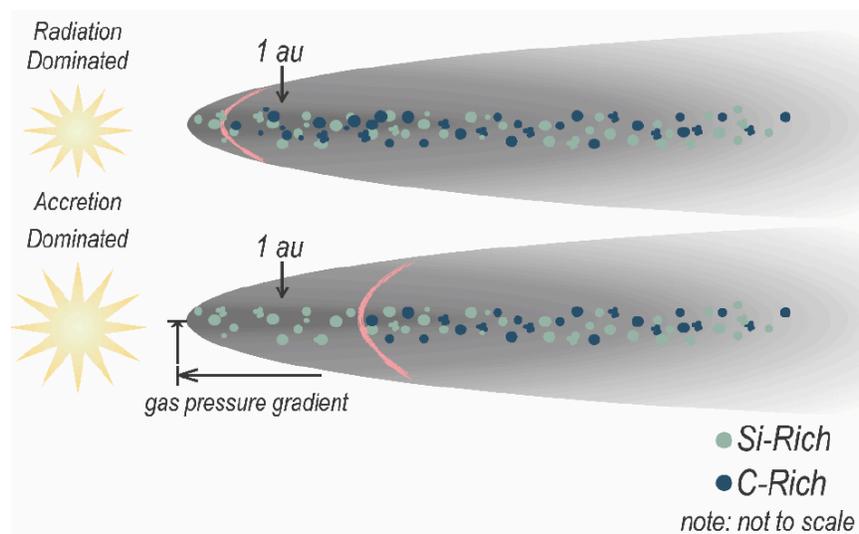

**Fig. 2. Schematic illustration of the soot line in a protoplanetary disk**. The soot line (red parabola) delineates the phase boundary between solid and gaseous carbon carriers. In the accretion-dominated disk phase, it is located far from the proto-Sun and divides carbon-poor dust and pebbles (green dots) from carbon-rich ones (dark blue dots). Within 1 Ma as a result of the transition to a radiation-dominated, or passive, disk phase, the soot line migrates inside of the Earth's current orbit. Note that the Si-rich and C-rich solids do not represent distinct reservoirs because carbonaceous material is likely associated with silicates. They are provided for ease in illustration.

A key proviso in the sublimation model is that the carbon-depleted precursor material of the inner solar system must have resided within the soot line. Our calculated thermal structures of disks illustrate that at accretion rates higher than 10⁻⁷ M☉ (solar mass)/yr the soot line resides beyond 1 au and potentially as far away as the asteroid belt (Fig. S3). If the primary materials accreted by the Earth are assembled during this phase, they would be carbon poor as the carbon grains would be destroyed while silicate grains could remain intact. For systems with accretion rates below ~10⁻⁷ M☉/yr, the soot line lies within the current

Earth orbit. Hence pre-planetary solids that exist at 1 AU during this phase are likely to be carbon-rich, similar to other objects known to form at low temperatures, such as comets Halley and 67/P. Observations show that the accretion rate in forming disk systems decreases so rapidly with time that the soot line would move inward to cross 1 AU within 1 Ma (9), thus the carbon-poor objects in Earth's chief feeding zone near 1 AU are likely present only during the first Ma of solar system history in the phase associated with high mass accretion rates. If the bulk carbon content of the Earth is low, then the majority of its source materials must have lost carbon through sublimation early in the nebula's history or by additional processes such as planetesimal differentiation. Constraining the fraction of carbon-depleted source material accreted by the Earth requires us to constrain the maximum amount of carbon in the bulk Earth.

In the near-surface reservoirs, carbon occurs in trace quantities at the level of hundreds of ppm by weight (10). The carbon content of the bulk silicate Earth (BSE) is more than three orders of magnitude below the solar composition (Fig. 1). The severe carbon deficit in the accessible part of the Earth has been attributed to its volatility and possible sequestration by the iron-rich core. The strong affinity of carbon for iron-rich alloys at low to moderate pressures suggests that the core could be the Earth's largest carbon reservoir (10-12). As such, it is helpful to seek geochemical and geophysical constraints on the carbon content of the core, which then places limits on carbon in the bulk Earth. Carbon is considered a candidate lighter element for the Earth's core because it may partially account for important physical properties, including density, sound speeds, elastic anisotropy, and the partially solidified state (11). A generous upper bound on core carbon content is obtained from density considerations (Fig. 3). The liquid outer core and solid inner core are less dense than pure iron at the relevant $P$-$T$ (pressure-temperature) conditions by 5-8 and 2-5%, respectively (11), and carbon could reduce or eliminate this density difference. With the equations-of-state of liquid iron and relevant iron-carbon alloys (13-15), we estimate that $5.0\pm0.6$ wt% and $4.5\pm0.5$ wt% carbon could account for the entire density deficits in both the liquid outer and solid inner core (16). Hence the amount of carbon in the core from the density constraints must be less than $5.0\pm0.5$ wt%. This estimate is undoubtedly generous, because the outer core contains significant amounts of sulfur, silicon, and oxygen (11) and possibly hydrogen (17).

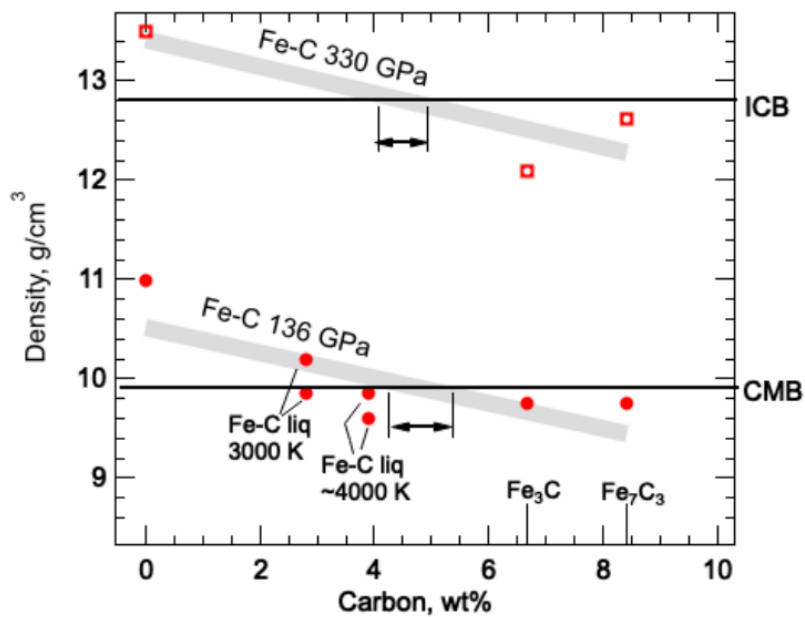

**Fig. 3  Generous upper bounds on carbon in the Earth's core, from density constraints**. The density of Fe-C alloy at the inner core boundary (ICB) or the core-mantle boundary (CMB) as a function of carbon content is estimated from that of iron and Fe-C alloys at 330 GPa and 5500±1500 K (red squares,*13*) and that at 136 GPa and 4500±1500 K (red filled circles,*14-15*), respectively. The thick trends are linear fits through experimental data on solid Fe-C alloys and compounds (upper), and liquid Fe-C alloys (lower). The horizontal black lines are the inferred densities at the ICB and CMB based on the Preliminary Reference Earth Model (PREM)(*16*). The horizontal arrows denote the ranges of carbon concentrations in iron-carbon alloys to match the core densities, considering uncertainties in core temperatures and mineral physics measurements. The maximum estimated carbon in the liquid outer and the solid inner core are 5.0±0.6 and 4.5±0.5 wt.%, assuming that carbon is the sole light element in the core to account for the observed densities (*16*).

More probable upper bounds may be obtained by comparing the sound velocities of Fe-C alloys with the values observed for the core. Existing data suggest that the presence of 1.0±0.6 wt.% carbon in liquid iron would match the compressional wave velocity ($V_p$) of the outer core (Fig. S5). Because this composition falls on the iron-rich side of the eutectic point of the Fe-C binary (Fig. S6), the solid inner core would contain less carbon and therefore is limited to <1.0±0.6 wt.%.

The maximum amount of carbon in the bulk Earth can then be calculated from estimated upper bounds on that of the bulk silicate Earth (BSE) and core. The BSE probably contains 140±40 ppm carbon by weight and most likely no more than 0.1 wt.% (*18*). With the core accounting for 32% of the Earth's mass, we arrive at a generous upper bound of 1.7±0.2 wt.% carbon, and a more probable upper bound of 0.4±0.2 wt.% (4000±2000 ppm by weight) carbon in the bulk Earth. We emphasize these are robust upper bounds and they are higher than the 530±210 ppm by weight estimate from geochemical constraints (*19*), and the recent estimate of 370-740 ppm from a multi-stage model of core formation using partition coefficients determined at relevant pressure and temperature conditions (*12*).

According to the sublimation sequence (Fig.1), carbon-rich source material is stable outside the soot line, but the abundance of carbon relative to Mg and CI falls precipitously from 4.74 to <0.47 due to sublimation of organic carbon carriers inside the soot line (Table S6). The estimated upper bound for the bulk Earth carbon at 0.2-1.9 wt.% corresponds to a maximum fraction of 1-7% carbon-rich source material in Earth's building blocks. The carbon-poor nature of the Earth implies that the majority of its precursor materials lost carbon. Because the soot line was already located inside of 1 AU by ~1 Ma, carbon loss from Earth's building blocks must occur very early in the solar history. If the sublimation temperature were higher than 500 K, then at a given disk cooling rate the soot line would reach 1 AU sooner. Consequently, the accretion of carbon-poor building blocks to the proto-Earth would have to take place earlier.

## Discussion

Earth's severe carbon deficit is consistent with the expectation of limited contribution of carbon-rich solids via pebble drift. The latest model for planet formation suggests that centimeter to meter-sized objects known as pebbles played a major role in delivering mass to growing bodies. Pebbles form via coagulation and settle to the dust-rich midplane where they are subject to forces leading towards an inward drift (*20*). These pebbles would thus carry both water and carbon from the outer solar system, beyond the soot line, to the inner solar system (*21*). Dynamic simulations show that pebbles drift inwards to the local pressure maximum (pressure bump) in the disk (*22-23*), which would nominally be the inner edge marking the destruction of the silicate dust (Fig. 2). However, a multi-Earth sized planet or giant-planet core in the disk would carve out a gap in the gaseous disk, with the outer edge representing a local pressure bump. Drifting pebbles would pile up there, thus diminishing the supply of carbon-rich precursors (*24*). Analysis of molybdenum and tungsten isotopes find evidence for two distinct reservoirs of meteorite parent bodies within the solar nebula that formed as early as 1 Ma and remained separate thereafter, presumably as a result of the formation of Jupiter's core (*25*). This scenario would reduce the supply of carbon to the inner solar system. It is worth noting that pressure bumps are pervasive in Ma-old disks (*22, 26*), and they may be induced via other means (*27*). Furthermore, in pebble accretion models, rapid formation of ~100 km sized bodies is further accelerated. If such bodies formed early in solar system history, especially within the first 0.1-0.2 Ma, radiogenic heating by short-lived $^{26}$Al can cause degassing (*28*). Thus, there are additional mechanisms of volatile loss through planetesimal degassing, also active in the first Ma of evolution, aside from the sublimation sequence discussed here that have caused the Earth's carbon deficit.

Very early carbon depletion in Earth's source material is supported by the carbon-poor nature of iron meteorites (*29*), which also formed in less than 1 Ma after CAIs (calcium- and aluminum-rich inclusions) (*30*). The notion of extensive carbon loss is also consistent with the composition of moderately or slightly volatile elements in the Earth (e.g., Na, K, Li, Si). The correlation between the bulk Earth abundance of an element and its half-mass condensation temperature (Fig. 1) implies thermally activated volatile loss, which requires that the majority of Earth's precursor materials were heated at least above the sublimation temperature of carbon-rich pre-solar grains. Sublimation is likely important for other life-essential elements such as nitrogen and hydrogen. At some point, the sequences of condensation and sublimation must merge (Fig. 1), and in the intermediate stages both processes could contribute to the creation and destruction of nebular solids, thereby setting the stage for the formation of habitable worlds.

## Materials and Methods
### Modeling soot line location as a function of time
We calculated sublimation temperatures of refractory carbon carriers using a kinetic rate law and experimentally constrained parameters (Fig. S1, Table S1). To determine the location of the soot line as a function of time, we calculate the thermal structure within a protoplanetary disk that is heated by internal viscous dissipation and irradiation from the central star (Fig. S3).

## Supplementary Materials
Supplementary text
Fig. S1-S6
Table S1-S6

## ACKNOWLEDGMENTS

We thank Larry Nittler and an anonymous reviewer for critically reading the manuscript and helping us improve and clarify**. Funding:** This research was supported by NSF grants AST 1344133, EAR 1763189, and AST1907653, and by the NASA Astrobiology Program, Grant NNX15AT33A. **Author contributions**: JL, EAB, GAB, FJC, MMH contributed equally to the project design and writing. **Competing interest:** The authors declare no competing interest. **Data and materials availability**: All data needed to evaluate the conclusions in the paper are present in the paper and/or the Supplementary Materials..

Supplementary Materials for

# Earth's carbon deficit caused by early loss through irreversible sublimation

J. Li*, E.A. Bergin, G.A.Blake,  F.J.Ciesla, M.M. Hirschmann

**\*Correspondence to: jackieli@umich.edu**

**This PDF file includes**
Supplementary text
Figs. S1-S6
Tables S1-S6

**Sublimation temperature from kinetic rate laws**

Sublimation is an irreversible process governed by kinetics, in contrast to the equilibrium process assumed in conventional condensation models. *Chyba* (*31*) proposed a kinetic rate law to describe the temperature ($T$) at which kerogen survives heating for a duration ($t$).

$T = E_a/\mathrm{R}[\ln(tA)^{-1}]$,

where R is the gas constant, $E_a$ is the activation energy of sublimation, and $A$ is a constant. This Arrhenius equation describes the anti-correlation between temperature and the lifetime of a carbon carrier (Fig. S1). Current theory suggests that during the early stages of proto-stellar evolution accretion events will likely occur in bursts of luminosity with an observationally constrained burst lifetime of order 100 years (*9*). This lifetime is therefore the characteristic heating timescale for the young inner regions of the disk. We calculated that the temperature to sublimate aliphatic hydrocarbon and kerogens as analogues for insoluble organic matter (IOM) in 100 years is 394 K and 478-508 K respectively, using experimentally constrained parameters (*31-32*, Table S1).

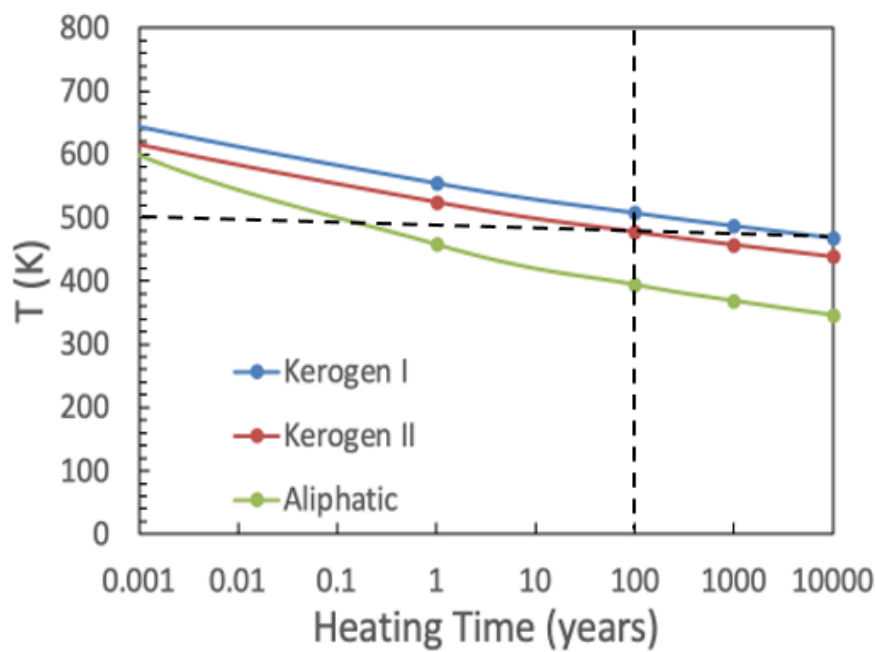

**Fig. S1. Kinetic rate law of carbon sublimation.** The solid blue, red, and green curves show the sublimation temperature as a function of the heating time during which type I kerogen and type II kerogen (*31*), and aliphatic organics (*32*), respectively, transform into gases. The dashed lines mark the 500 K sublimation temperature for a characteristic heating time of 100 years.

**Table S1 Kinetic rate law parameters and calculated sublimation Temperature**

|  | Kerogen Type I (*31*) | Kerogen Type II (*31*) | Aliphatic organics (*32*) |
|---|---|---|---|
| $E_a$ (kJ/mol) | 2.31E+05 | 2.04E+05 | 1.09E+05 |
| $A$ (s$^{-1}$) | 1.70E+14 | 6.70E+12 | 8.70E+04 |
| $T$ (K)* | 508 | 478 | 394 |

*Sublimation temperature to shed carbon in 100 years

To further examine thermal destruction of IOM in chondrites we plotted the compositional information from Alexander (*33*) as a function of the inferred temperatures from Cody et al. (*34*) in Fig. S2. This exercise is problematic for several reasons, including (a) the data are rather scattered, (b) the normalized decay is very strongly dependent on the assumed initial carbon concentration, (c) the carbon loss is probably a result of thermal or aqueous processing on chondrites, and does not tell us what happens in the dust or in pebbles, and (d) it is not a good assumption that all of these bodies accreted from material with the same initial carbon concentration. Nevertheless, the Alexander-Cody plot shows that upon heating to 300-500 K, chondrites lose 60-70% of its IOM (Fig. S2). This result supports the effective sublimation temperatures of IOM calculated from kinetic rate laws (Table S1).

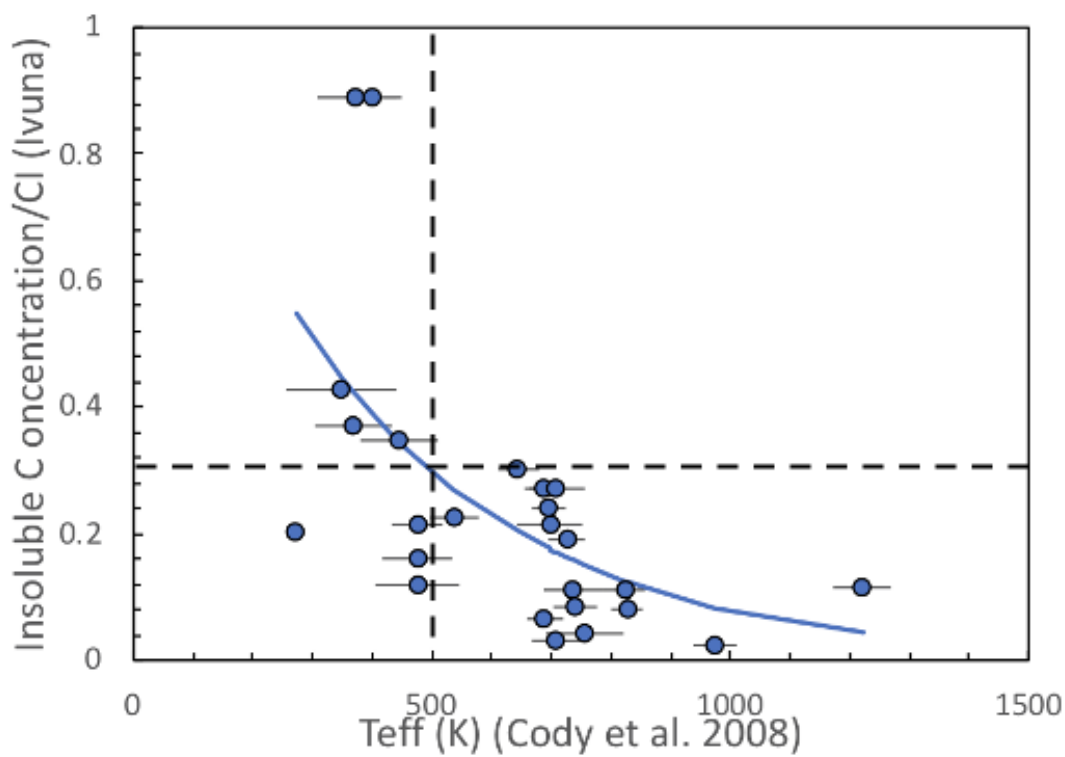

**Fig. S2 Amount of insoluble carbon in chondrites as a function of temperature.** Insoluble carbon in chondrites, normalized to CI Ivuna with 2.25 wt.% insoluble carbon (*33*) versus Teff, the "Organic Temperature" inferred from the abundance of graphite-like carbon (*34*). Two chondrites, Orgueil (CI) and Tagish Lake (T2) have 2 wt.% C, whilst the others all have no more than half that. The solid blue curve is a fit to a simple exponential function. The dashed lines mark 70% loss of IOM at temperatures below 500 K.

The calculated sublimation temperatures are in general agreement with literature values based on heating experiments (e.g., *4, 35,* Table S2). A number of pyrolysis studies found that some IOM may survive heating in the laboratory at temperature above 500 K in time scales ranging from near-instantaneous to several hours (e.g., *36-38*). However, these studies were not designed to investigate the kinetics of thermal decomposition. Refractory organics may survive for hours or days in a lab, but not weeks to centuries in a nebular disk.

Considering that the refractory organics sublimate in 100 years at 478-508 K (Table S2), we adopt 500 K as the effective sublimation temperature of refractory carbon carriers in the solar nebula. This kinetics-based sublimation temperatures capture the irreversible nature of the transformation of refractory carbon carriers into gases and is used to define the soot line in the disk (Fig. 1 and Fig. S3).

**Table S2 Sublimation temperatures of carbon carriers in the solar nebula**

| Carrier | Sublimation $T$, K | Fraction of solar carbon (Table S5) |
|---|---|---|
| CO | 35 (*3*) | 0.21 |
| $CH_4$ | 40 (*3*) | 0.0 |
| Other solids | 50 (assumed) | 0.05 |
| $CO_2$ | 80 (*3*) | 0.26 |
| $CH_3OH$ | 120 (*3*) | 0.027 |
| Aliphatic organics | 394 (Table S1) | 0.05 |
| Aromatic organics | 425 (*4*) | 0.05 |
| Refractory organics | 478-508 (Table S1) | 0.30 |
| Amorphous C | 1100 (*4*) | 0.05 |
| SiC | >1100 (*51*) | 2.50E-06 |
| Total | | 0.997 |

**Disk Modeling**

To determine the location of the soot line as a function of time, we calculate the thermal structure within a protoplanetary disk that is heated by internal viscous dissipation and irradiation from the central star. The viscous dissipation is calculated assuming a mass accretion rate as determined by the analytic formula derived by Manara et al. (*39*). The median values are plotted within an envelope representing 0.5 dex ranges given the scatter in the observations (Fig. S3). As the mechanisms for driving the transport of mass in protoplanetary disks remains uncertain, we follow the classical alpha-disk model of Shakura and Sunyaev (*40*) to describe this transport arising from internal viscosity within the gas, adopting alpha=$10^{-3}$, a value consistent with astronomical constraints (*22, 41*). The thermal profile is calculated assuming an opacity of 1 $cm^2$/g at all times. The lower bound on the soot line location is represented by the irradiated profile (Fig. S3), where viscous evolution is negligible. For systems with accretion rates below ~$10^{-7}$ $M_*$/yr, the midplane temperature due to irradiation only is estimated following the method described in (*22*), using the following expression:

$$T_d(r) = \frac{\frac{1}{2}\phi L_*}{4\pi r^2 \sigma_{SB}}$$

where $T_d(r)$ is the midplane dust or solid temperature as a function of radius $r$, $\phi$ is the flaring angle, $L_*$ is the stellar luminosity, and $\sigma_{SB}$ is the Stefan-Boltzmann constant. Current models of disk observations assume $\phi$=0.02 (*22*). For a young solar-mass star on the pre-main sequence, its luminosity at the relevant age of a few Ma is ~2 $L_*$ (*42*).

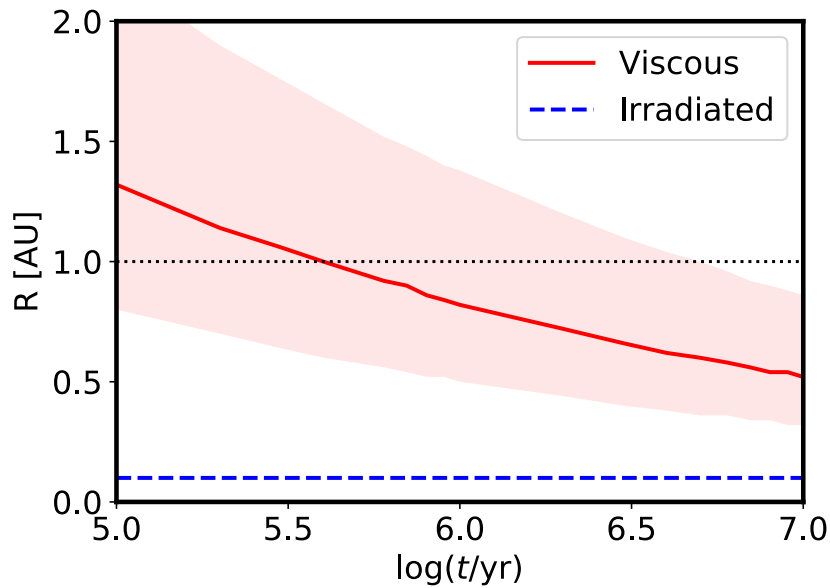

**Fig. S3. Migration of the soot line at 500 K with time in a representative protoplanetary disk.** The red solid curve traces the location of the soot line moving toward the central star with time in an accretion-dominated disk with a median value of alpha at $10^{-3}$ within an envelope representing 0.5 dex ranges. The blue dashed line marks the location of the soot line at 0.1 AU in an irradiated disk of a solar-mass star. The black dotted line denotes 1 AU.

We stress that these calculations are conservative in the sense that the calculated disk temperatures are likely too hot. The parameter alpha corresponds to the stress to pressure ratio and controls the rate of internal dissipation. In general, as the gaseous disk dissipates and grains evolve, $\alpha$ will change and the disk cools more efficiently. This is not accounted for in our calculations. The opacity parameter sets how readily heat can diffuse through the disk and get lost to the surroundings. Small grains are the primary source of opacity in disks. As they grow into pebbles and planetesimals over time, the opacity will decrease, making the disks cooler than calculated here. Thus the location of the soot line in the viscous model should be taken as an upper limit, particularly at later times. At t >1 Ma, our simple approach likely overestimates the expected temperatures for an evolving disk and the radiation dominated limit is likely approached.

**Bulk silicate Earth composition**

The compositions of the bulk silicate Earth (BSE) and CI show striking similarities and obvious differences. Comparison with CI reveals that the relative abundances of refractory elements in the BSE are chondritic but volatile elements are depleted with respective to CI, implying that that of the Earth accreted more volatile-poor source materials and/or loss more volatiles through additional processes. To first order, the condensation sequence works for elements that are more volatile than C, but fails to account of the forms and abundances of C in meteoritic and cometary records.

In Fig. 1, an element's relative abundance to Mg and CI is calculated as the ratio of the relative abundance of an element to Mg in BSE to its relative abundance to Mg in CI, so that the relative abundance of Mg, a major element in both the BSE and CI that divides refractory elements and volatile elements, is anchored at 1 (Table S3).

**Table S3 Element abundances in bulk silicate Earth**

|     | BSE (2) wt.% | BSE relative to Mg | CI (2) wt.% | CI relative to Mg | BSE relative to Mg and CI[*] |
| --- | --- | --- | --- | --- | --- |
| O | 44 | 1.93 | 48.2 | 4.99 | 0.39 |
| Mg | 22.8 | 1.00 | 9.65 | 1.00 | 1.00 |
| Si | 21 | 0.92 | 10.65 | 1.10 | 0.83 |
| Fe | 6.26 | 0.27 | 18.1 | 1.88 | 0.15 |
| Ca | 2.53 | 0.11 | 0.92 | 0.10 | 1.16 |
| Al | 2.35 | 0.10 | 0.86 | 0.09 | 1.16 |
| Ni | 0.196 | 0.01 | 1.05 | 0.11 | 0.08 |
| S | 0.025 | 1.10E-03 | 5.4 | 0.56 | 2.0E-03 |
| C | 0.014 | 6.14E-04 | 3.5 | 0.36 | 1.7E-03 |
| Total | 99.175 | | 98.33 | | |

[*]Calculated as the ratio of an element's relative abundance to Mg in BSE to that in CI.

## Carbon carriers in the solar nebula

Solid carbon carriers in the solar nebula include ices and refractory phases. The abundance of total carbon in nebular solid is calculated from the solar system composition excluding H and He (Table S4). For comparison with the bulk silicate Earth and Earth, the relative abundance to Mg and CI is used in Fig. 1.

**Table S4 Element abundances in the solar system**

|     | Solar (1) log A(El) | Solar solid[*] wt.% | Solar solid relative to Mg | CI (2) wt.% | CI relative to Mg | Solar solid relative to Mg and CI[#] |
| --- | --- | --- | --- | --- | --- | --- |
| H | 12 | | | | | |
| He | 10.98 | | | | | |
| C | 8.46 | 16.8 | 3.44 | 3.5 | 0.36 | 9.48 |
| N | 7.9 | 5.4 | 1.10 | | | |
| O | 8.76 | 44.8 | 9.15 | | | |
| Ne | 7.95 | 8.7 | 1.77 | | | |
| Mg | 7.62 | 4.9 | 0.99 | 9.65 | 1.00 | 0.99 |
| Al | 6.54 | 0.5 | 0.09 | 0.86 | 0.09 | 1.04 |
| Si | 7.61 | 5.6 | 1.13 | 10.65 | 1.10 | 1.03 |
| S | 7.26 | 2.8 | 0.58 | 5.4 | 0.56 | 1.03 |
| Ca | 6.41 | 0.5 | 0.10 | 0.925 | 0.10 | 1.07 |
| Fe | 7.54 | 9.5 | 1.93 | 18.1 | 1.88 | 1.03 |
| Ni | 6.29 | 0.6 | 0.11 | 1.05 | 0.11 | 1.05 |

[*]Solar system composition excluding H and He
[#]Calculated as the ratio of an element's relative abundance to Mg in a solar solid to that in CI.

In Table S5 we provide a list of solid-state carbon carriers (4, 43), assuming that 50% of the solar carbon takes refractory forms and 50% would be in the ices at temperatures below 125 K (3). In the sublimation sequence, upon heating CO will be the first to be lost to the gas, whereas in the condensation sequence, upon cooling CO will be the last to transform from gas to solid (Fig. S4). Refractory organics are likely pre-solar (*5*), although some could be created through reactions in the solar nebula (*44-45*). They are not products of condensation from the solar nebula (*1, 3-4*).

**Table S5 Carbon carriers in solar nebula**

|  | Fraction of C in ices | Fraction of solar C[*] | Relative Abundance to Mg (Table S4) | Relative abundance to Mg and CI (Table S4) |
|---|---|---|---|---|
| *Öberg et al. 2011* (*43*) |  |  |  |  |
| $CO_2$ | 0.521 | 0.26 | 0.895 | 2.47 |
| CO | 0.425 | 0.212 | 0.730 | 2.01 |
| $CH_4$ | 0.0 | 0.0 | 0.000 | 0.00 |
| $CH_3OH$ | 0.055 | 0.027 | 0.094 | 0.26 |
|  | Fraction of C in refractory solids |  |  |  |
| *Gail and Treiloff 2017* (*4*) |  |  |  |  |
| Refractory organics | 0.6 | 0.3 | 1.032 | 2.85 |
| Aromatic organics | 0.1 | 0.05 | 0.172 | 0.47 |
| Aliphatic organics | 0.1 | 0.05 | 0.172 | 0.47 |
| Amorphous C | 0.1 | 0.05 | 0.172 | 0.47 |
| Others | 0.1 | 0.05 | 0.172 | 0.47 |
| *Alexander et al. 2017* (*51*) |  |  |  |  |
| SiC | 5.00E-06 | 2.50E-06 | 8.60E-06 | 2.37E-05 |

[*]Assuming 50% in ices, 50% in refractory solids (*3*).

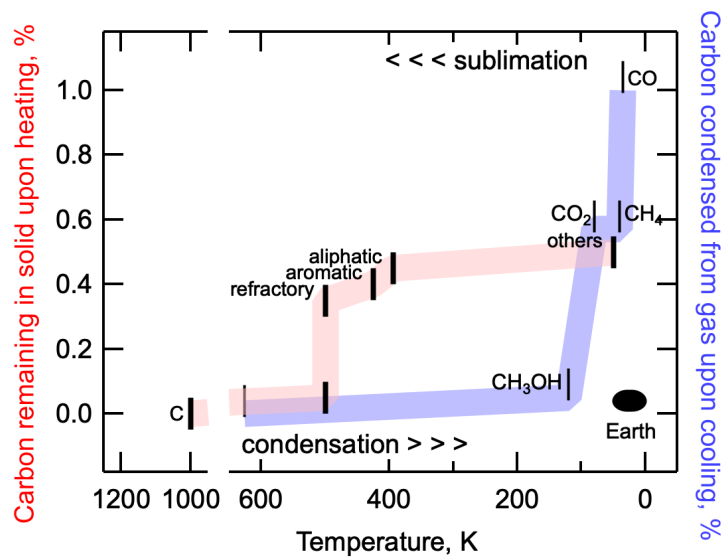

**Fig. S4 Sublimation and condensation sequences of carbon in the solar nebula**. Upon heating (red trace) the fraction of carbon remaining in dust decreases as ices sublimate at temperatures below 120 K and most refractory carbon carriers (thick vertical bars) in the inherited ISM grains sublimate near a characteristic temperature of ~500 K in 100 years. Upon cooling (blue trace) the remaining fraction of carbon in gas increases but >90% of the gaseous carbon carriers (thin vertical bars) do not condense until the temperature falls below 100 K. The oval in the lower right corner represents the estimated upper bound for the bulk Earth carbon at 0.2-1.9 wt.%, corresponding to 1-7% of the total carbon available in the solar nebula. Data are from Table S2, Table S4, and Table S6.

In a more recent model (*46*), the ISM diffuse cloud dust contains 83 ppm hydrocarbons and 40 ppm PHAs, corresponding to a smaller fraction of aliphatic organics than adopted here (*4, 47*). Because aliphatics are more volatile and sublimate at relatively low temperature, the differences do not affect our conclusion. Gail and Treiloff (*4*) isolate a fraction of carbon as highly refractory pure carbon phases (amorphous carbon, graphite, and/or nano-diamonds) with sublimation temperatures greater than 1000 K. This fraction is inferred from an analysis of Comet Halley dust (*48*) and not meteoritic material that survived in the inner solar system such as CI chondrites. In Fig. 1, we do not include this fraction in our sublimation sequence for the inner solar system because the carriers of this signature are amongst the lightest and smallest particles (*4, 48*). At sizes of <0.05 μm, these small particles are readily destroyed in the inner solar system via oxidation near the surface of the disk (*49-50*). Instead, we assume that the highly refractor fraction of C is of order 0.1 x CI with a tail to lower values as temperature increases (Fig. S2). At a baseline level the minimum available amount of highly refractory carbon in the inner solar system is carried by SiC at 5 ppm as traced by meteoritic material (*51*). We note that possible survival of highly refractory carbon at 1100 K would strengthen our conclusion that the Earth's source material lost carbon early in the solar history, when the temperature near 1 AU was sufficiently high to remove this component.

**Additional constraints on the maximum carbon content of Earth's core**

**From sound velocities and phase relation**: Existing data suggest that the presence of 1.0±0.6 wt.% carbon in liquid iron increases the compressional wave velocity ($V_p$) of liquid iron to match the observed value of the outer core (Fig. S5), whereas $Fe_7C_3$ containing as much as 8.4 wt.% carbon can reproduce the anomalously low shear wave velocity ($V_s$) and liquid-like high Poisson's ratio of the inner core (*52-53*). With 1.0±0.6 wt.% carbon in the outer core and 8.4 w% carbon in the inner core, we obtain an upper bound of 1.3±0.6 wt.% carbon in the core.

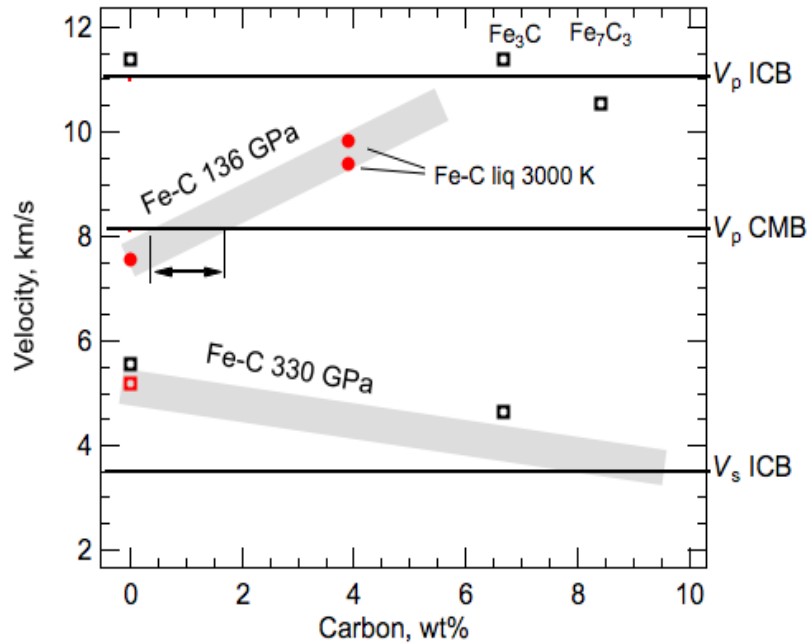

**Fig. S5  Upper bound on the carbon content of Earth's core from velocity constraints.** The maximum amount of carbon in the liquid outer core is estimated at 1.0±0.6 wt.% (horizontal arrow), assuming that carbon is the only light element responsible for the difference between the sound velocities of iron and that of the core, marked by the horizontal black lines (*16*). The $V_p$ of Fe-C alloy at the inner core boundary (ICB), $V_p$ of Fe-C alloy at the core-mantle boundary (CMB), and $V_s$ of Fe-C alloy at the ICB as a function of carbon content are estimated from that of iron and Fe-C alloys at 330 GPa and 300 K (open black squares along the gray line, *16*), that at 136 GPa and 4500±1500 K (red filled circles along the upper thick gray line, *14-15*) and that at 330 GPa and 5500±1500 K (red open squares along the lower thick gray line), respectively.

With 1.0±0.6 wt.% carbon, outer core composition would fall on the iron-rich side of the eutectic point in the Fe-C binary system (*53-55*, Fig. S6). As a result, an iron-carbon alloy, instead of Fe₇C₃, would be the solid phase to form the inner core. This scenario is supported by studies that found iron carbide too light or too fast for the inner core, and estimated that the inner core contains 1.5 wt.% carbon (*56-59*). In this case, the inner core composition would contain less carbon than the outer core, and therefore the carbon content of the inner core is limited to 1.0±0.6 wt.%. Accordingly, the upper bound for the carbon content of the core is reduced to 1.0±0.6 wt%.

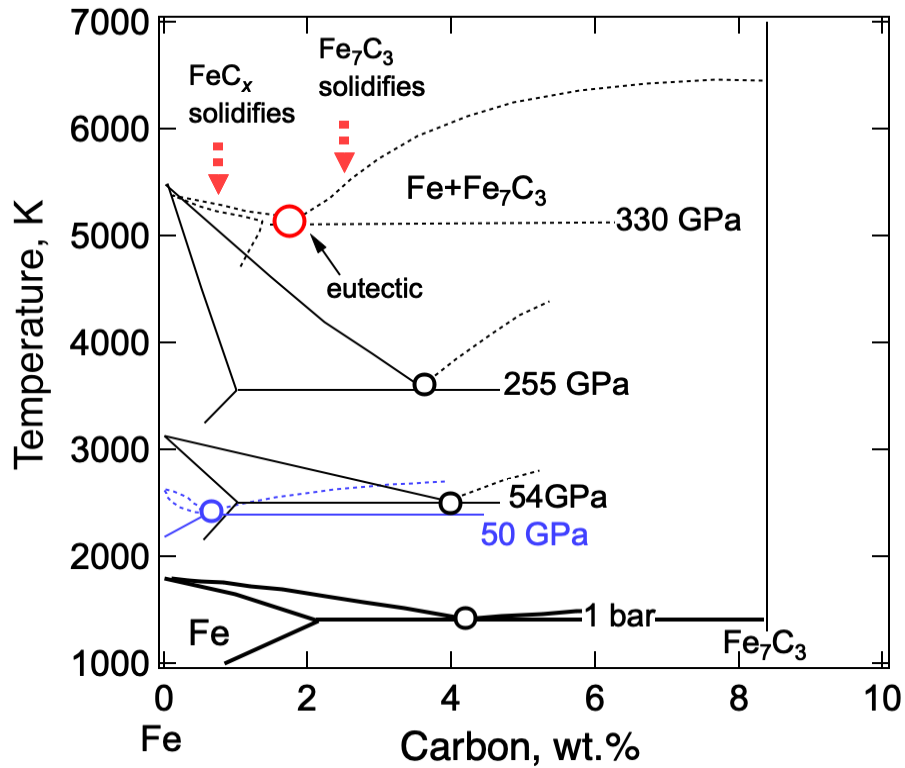

**Fig. S6 Constraints on the maximum carbon content of Earth's core from iron-carbon binary phase relation.** Existing results at 54-255 GPa (*53*) and 330 GPa (*54*) ), with the exception of one result at 50 GPa (Blue, *55*), suggest that the eutectic composition of the Fe-C binary system contain more than 2 wt.% carbon, and therefore an Fe-C alloy, instead of $Fe_7C_3$, would solidify and form the inner core.

**From metal-silicate partitioning:** Once accreted to the Earth, the distribution of carbon depends on its partitioning between the iron-rich core and the bulk silicate Earth including the fluid envelope (atmosphere and hydrosphere). The strong affinity of carbon for iron-rich alloys at relatively low pressures suggest that sequestration in the core could be partially responsible for the depletion of carbon in the silicate Earth relative to primitive chondrites (*e.g., 60*), although recent experiments suggest that carbon may become much less siderophile or even lithophile at pressures and temperatures that are relevant to core formation in a deep magma ocean (*12 Fischer*). Using updated parametrization for the partition coefficient of carbon between metal and silicate, Fischer et al. (*12*) obtained an upper bound of 0.1-0.2 wt% in the core, and 370-740 ppm carbon in the bulk Earth. This bulk Earth carbon content overlaps with previous estimates 530±210 ppm for the bulk Earth, derived from an estimate of 765±300 ppm in the bulk silicate Earth and assuming a carbon-free core (*19*). Both values may be viewed as conservative upper bounds and require even larger fractions of carbon-poor building blocks for the Earth.

**Table S6 Upper bounds on the carbon content of bulk Earth**

|   | Bulk Silicate Earth (*2*) | Bulk Earth upper bound generous | Bulk Earth upper bound probable | CI (*2*) | C-rich Source material[*] |
|---|---|---|---|---|---|
| C | 140±40 ppmw | 1.7±0.2 wt.% | 0.4±0.2 wt.% | 3.5 wt.% | |

| average | 140 | 1.7 | 0.4 | | |
| + | 180 | 1.5 | 0.2 | | |
| - | 100 | 1.9 | 0.6 | | |

Relative abundance to Mg: C/Mg

| average | 6.14E-04 | 0.11 | 0.03 | 0.36 | 1.73 |
| + | 7.89E-04 | 0.10 | 0.01 | | |
| - | 4.39E-04 | 0.12 | 0.04 | | |

Relative abundance to Mg and CI: (C/Mg) / (C/Mg in CI)

| average | 1.71E-03 | 0.30 | 0.07 | | 4.74 |
| + | 2.19E-03 | 0.27 | 0.04 | | |
| - | 1.22E-03 | 0.34 | 0.11 | | |

Fraction of C-rich source material in bulk Earth

| average | | 0.06 | 0.01 | | |
| + | | 0.06 | 0.01 | | |
| - | | 0.07 | 0.02 | | |

| Mg, wt.% | 22.8 | 15.4 | | 9.65 | |

[*]After C-carrying ices sublimate, the C/Mg ratio is reduced to half of the initial value (Table S2).

**Glossary**

Aliphatic hydrocarbon: Hydrocarbons based on chains of C atoms, includes alkanes such as methane $CH_4$, alkenes such as ethene $C_2H_4$, and alkynes such as acetylene $C_2H_2$.

Aromatic hydrocarbon: Hydrocarbons containing one or more six-carbon ring as in benzene $C_6H_6$.

CHON: A mnemonic acronym for the four most common elements in living organisms: carbon, hydrogen, oxygen, and nitrogen. Note that CHON particles refer to kerogen.

Hydrocarbon: Organic chemical compound composed exclusively of hydrogen and carbon atoms such as methane $CH_4$.

IDP: Interplanetary dust particle, also called micrometeoroid, micrometeorite, or cosmic dust particle, a small grain, generally less than a few hundred micrometers in size and composed of silicate minerals and glassy nodules but sometimes including sulfides, metals, other minerals, and carbonaceous material, in orbit around the Sun.

IOM: Insoluble organic matter. It is major constituent of organic matter in extraterrestrial materials. Includes both the more refractory macromolecular carbon (=CHON=kerogen) as well as the less refractory aromatic and aliphatic hydrocarbons.

ISM: Interstellar medium. the matter and radiation that exists in the space between the star systems in a galaxy. This matter includes gas in ionic, atomic, and molecular form, as well as dust and cosmic rays. It fills interstellar space and blends smoothly into the surrounding intergalactic space.

Kerogen: A solvent-insoluble organic matter in terrestrial rocks. Mainly consists of paraffin hydrocarbon, also called alkane. Has the general formula $C_nH_{2n+2}$. A terrestrial analog to IOM in meteorites.

Lithophile element: Lithophile means rock-loving. A lithophile element readily combines with oxygen and therefore prefers silicate-rich mantle to iron-rich core. Examples include Mg, Si, Ca, Al, and K.

Macromolecular organics: Including PAHs and insoluble organics matter.

PAH: Polycyclic aromatic hydrocarbon, such as naphthalene $C_{10}H_8$.

Pyrolysis: Thermal Decomposition of materials at high temperatures in an inert atmosphere.

ppm: Parts per million. The number of units of mass of element per million units of total mass

Refractory organics: Predominantly hydrocarbon in nature, with the carbon distributed between the aromatic and aliphatic forms.

Siderophile element: Siderophile means iron-loving. A siderophile element readily alloys with iron and therefore prefers iron-rich core to silicate-rich mantle. Examples include Ni, Au, Pt, S, and C.